\documentclass[a4paper,fleqn,usenatbib]{mnras}

\usepackage{newtxtext,newtxmath}

\usepackage[T1]{fontenc}
\usepackage{ae,aecompl}

\usepackage{graphicx}	
\usepackage{amsmath}	
\usepackage{amssymb}	
\usepackage{xspace}

\setcitestyle{notesep={}} 

\newcommand{\dcc}{LIGO-P1800175\xspace}
\newcommand{\sw}[1]{\texttt{#1}} 
\newcommand{\gw}[1][]{gravitational wave#1 (GW#1)\renewcommand{\gw}[1][]{GW##1\xspace}\xspace}
\newcommand{\ns}[1][]{NS#1\xspace}
\newcommand{\dns}[1][]{Double Neutron Star#1 (DNS#1)\renewcommand{\dns}[1][]{DNS##1\xspace}\xspace}
\newcommand{\bns}[1][]{binary neutron star#1 (BNS#1)\renewcommand{\bns}[1][]{BNS##1\xspace}\xspace}
\newcommand{\eos}[1][]{equation#1 of state (EoS)\renewcommand{\eos}[1][]{EoS\xspace}\xspace}
\newcommand{\gmm}[1][]{Gaussian mixture model#1 (GMM#1)\renewcommand{\gmm}[1][]{GMM##1\xspace}\xspace}
\newcommand{\skgmm}{\sw{sklearnGMM}\xspace}
\newcommand{\xdgmm}{\sw{XDGMM}\xspace}
\newcommand{\lmxb}[1][]{Low Mass X-ray Binary#1 (LMXB#1)\renewcommand{\lmxb}[1][]{LMXB##1\xspace}\xspace}
\newcommand{\cv}{cross-validation (CV)\renewcommand{\cv}{CV\xspace}\xspace}
\newcommand{\numdns}{15\xspace}
\newcommand{\cpnest}{\sw{CPNest}\xspace}
\newcommand{\python}{\sw{python}\xspace}
\newcommand{\mixinf}{\sw{MixtureInf}\xspace}
\newcommand{\citehuang}{\citet{Huang:2018ese}\xspace}
\newcommand{\huangshort}{Huang+\xspace}
\newcommand{\huang}{\citehuang\renewcommand{\huang}{\huangshort}}
\newcommand{\Mtot}{M_{\,\mathrm{T}}}
\newcommand{\Msun}{M_\odot}
\newcommand{\Nbins}{N_\mathrm{bins}}

\newcommand{\AICc}{\mathrm{AICc}}
\newcommand{\BIC}{\mathrm{BIC}}

\newcommand{\Ndata}{N_{\mathrm{data}}}
\newcommand{\Ncoef}{N_{\mathrm{coeffs}}}
\newcommand{\likeli}{\mathcal{L}}
\newcommand{\logL}{\ln\likeli}
\newcommand{\logtenL}{\log_{10}\likeli}
\newcommand{\Ncomp}{N_{\mathrm{comp}}}
\newcommand{\fspin}{f_{\mathrm{spin}}}
\newcommand{\Hyp}{\mathcal{H}}
\newcommand{\pInfo}{\mathcal{I}}
\newcommand{\prob}[2]{P\left(#1\,|\,#2\right)}
\newcommand{\evi}{\mathcal{Z}}
\newcommand{\logevi}{\ln\mathcal{Z}}
\newcommand{\Nlive}{N_{\,\mathrm{live}}}

\title[DNS model selection]{Galactic Double Neutron Star total masses\\ and Gaussian mixture model selection}

\author[D.~Keitel]{David Keitel$^{1,2}$\thanks{E-mail: david.keitel@ligo.org}
\\
$^{1}$University of Glasgow, School of Physics and Astronomy, Kelvin Building, Glasgow G12 8QQ, Scotland, United Kingdom \\
$^{2}$University of Portsmouth, Institute of Cosmology and Gravitation, Portsmouth PO1 3FX, United Kingdom
\vspace{-\baselineskip}
}

\date{\dcc [draft version: 14 January 2019]} 

\pubyear{2019}

\begin{document}
\label{firstpage}
\pagerange{\pageref{firstpage}--\pageref{lastpage}}
\maketitle

\begin{abstract}
\citehuang
have analysed the population of 15 known galactic \dns[s]
regarding the total masses of these systems.
They suggest the existence of two sub-populations,
and report likelihood-based preference
for a two-component Gaussian mixture model
over a single Gaussian distribution.
This note offers a cautionary perspective on model selection for this data set:
Especially for such a small sample size,
a pure likelihood ratio test
can encourage overfitting.
This can be avoided by penalising models
with a higher number of free parameters.
Re-examining the \dns total mass data set
within the class of Gaussian mixture models,
this can be achieved through several simple and well-established statistical tests,
including information criteria (AICc, BIC),
cross-validation,
Bayesian evidence ratios
and a penalised EM-test.
While this re-analysis confirms the basic finding that a two-component mixture
is consistent with the data,
the model selection criteria
consistently indicate that
there is no robust preference
for it over a single-component fit.
Additional \dns discoveries will be needed
to settle the question of sub-populations.
\end{abstract}




\vspace*{-5\baselineskip}

\section{Introduction}

The population of galactic \dns[s] --
or \bns[s], as the \gw community prefers to call them --
is of high interest
as a locally accessible predictor
for the population of merging binaries in the wider Universe,
which has recently become accessible to \gw observations
with LIGO and Virgo~\citep{TheLIGOScientific:2017qsa}.
Traditionally, a lot of work has focused on using the observed galactic sample
to predict coalescence rates \citep[see][ and references therein]{Aasi:2013wya},
though the distribution of component masses
has also been studied~\citep{Schwab:2010jm,Zhang:2010qr,Ozel:2012ax,Kiziltan:2013oja}.

In a recent paper, \huang (in the following: \huangshort)
have considered the total gravitational masses $\Mtot$
of \numdns known \dns[s].
$\Mtot$ is of special interest in predicting
the fate of binary merger remnants
and for studies of the nuclear \eos~\citep{Baiotti:2016qnr,Margalit:2017dij,Ma:2017yva,Abbott:2017dke,Abbott:2018hgk,Abbott:2018exr}.
\huang point out an apparent bimodality in the distribution of $\Mtot$,
and with the help of \gmm[s] and a likelihood ratio test,
they arrived at a $2\sigma$ preference for two components over one.

In this note,
I suggest additional statistical tests
not originally considered by \huang,
and caution against relying on
likelihood-ratio tests alone,
especially when applied to small data sets.
Hence, let us re-evaluate
the suggested preference for a two-component \gmm
fit to the observed \dns $\Mtot$ distribution
with a series of simple tests.
First, for completeness,
(i) visual inspection of the data set
(Sec.~\ref{sec:data})
and
(ii) \gmm fitting and likelihood-ratio tests
(Sec.~\ref{sec:gmms})
are briefly summarised.
The additional hypothesis test methods include
(iii) information criteria
(AICc and BIC)
that penalise underconstrained parameters
(Sec.~\ref{sec:aicbic}),
(iv) a cross-validation test
to understand the impact of individual \dns systems on
model selection
(Sec.~\ref{sec:crossval}),
(v) Bayesian evidence computation through nested sampling
(Sec.~\ref{sec:nestsamp}),
and (vi) a penalised EM-test
(Sec.~\ref{sec:em-test}).

To provide more context for the model selection results,
the same criteria are also applied on additional examples:
simulated larger $\Mtot$ data sets
(appendix~\ref{sec:appendix-sims})
and a physically different,
but statistically not dissimilar data set
of \ns spins from~\citet{Patruno:2017oum}
(appendix~\ref{sec:appendix-lmxbs}).

\vspace{-0.5\baselineskip}

\section{GMM model selection on the DNS mass distribution}
\label{sec:tests}

\subsection{Data set and visual inspection}
\label{sec:data}

This analysis reuses the $\Mtot$ values,
with measurement errors,
as collected in Table~I of~\huang.
(For references to the original measurements,
please see that table.)
Individual component masses
(pulsars and companions)
are not considered here;
this could be a fruitful topic for further study.

\begin{figure}
\includegraphics[width=\columnwidth]{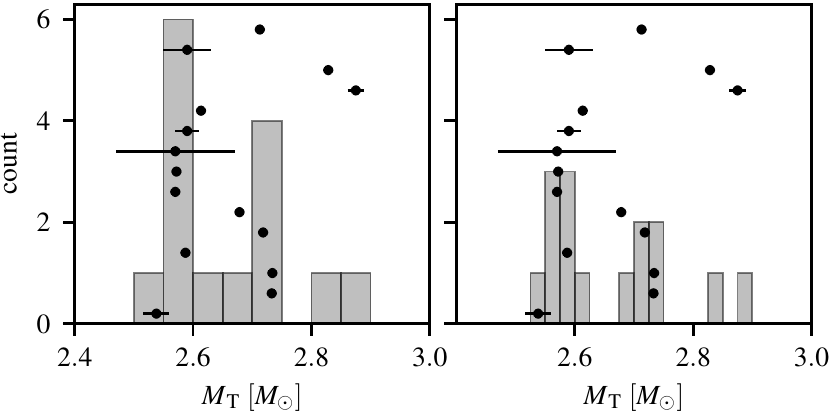}
 \caption{The data set of total masses $\Mtot$ from Table~I of~\huang.
          The left panel shows a histogram with the same number of bins ($\Nbins=12$)
          between $2.4$ and $3.0\,\Msun$
          as in Fig.~1 of~\huang.
          The right panel uses twice as many bins.
          The $\Mtot$ values with error bars for individual \dns[s]
          are overlaid as scatter plots;
          the vertical placement is just for ease of viewing
          and has no numerical meaning.
          \vspace{-\baselineskip}
         }
 \label{fig:binning}
\end{figure}

Histograms of the $\Mtot$ data set are shown in Fig.~\ref{fig:binning}.
It compares the original binning from~\huang
with the alternative choice of twice as many bins.
Visually, $\Nbins=12$ makes a two-component fit appealing,
while $\Nbins=24$ might even tempt the viewer into fitting three components.
Note that the total number of data points is only \numdns.
The overlaid scatter plots illustrate
the large range in error bar magnitudes
on the $\Mtot$ measurements,
and a concentration
of more uncertain measurements
near the apparent `lower peak'.

\subsection{GMMs and likelihood ratios}
\label{sec:gmms}

For $\Ndata$ data points $x_n$,
the basic likelihood function for a \gmm
with $\Ncomp$
means $\mu_k$,
widths $\sigma_k$
and component weights \mbox{$C_k \in [0,1]$}
is the product
$\prod_{x_n} \likeli(x_n|\,\{\mu_k,\sigma_k,C_k\})$
of
\begin{equation}
 \label{eq:gmm}
 \likeli \left( x_n\,|\,\{\mu_k,\sigma_k,C_k\} \right)
 = \sum\limits_{k=1}^{\Ncomp} \frac{C_k}{\sqrt{2\pi}\,\sigma_k} \,
   \exp\left( -\frac{\left(x_n-\mu_k\right)^2}{2\sigma_k^2} \right) \,.
\end{equation}
This can be amended to include measurement errors
by assuming each $x_n$ to come
from a Gaussian with mean $\mu_n$ and width $\sigma_n$,
then marginalising over $x_n$ as nuisance variables:
\begin{align}
 \label{eq:gmm_errs}
 \likeli & \left( \mu_n,\sigma_n\,|\,\{\mu_k,\sigma_k,C_k\} \right) \nonumber \\
 =& \int\limits_{-\infty}^{\infty}
    \frac{\mathrm{d}x_n}{\sqrt{2\pi}\,\sigma_n} \,
    \exp\left( -\frac{\left(x_n-\mu_n\right)^2}{2\sigma_n^2} \right)
    \sum\limits_{k=1}^{\Ncomp} \frac{C_k}{\sqrt{2\pi}\,\sigma_k} \,
    \exp\left( -\frac{\left(x_n-\mu_k\right)^2}{2\sigma_k^2} \right) \nonumber \\
 =& \sum\limits_{k=1}^{\Ncomp} \frac{C_k}{\sqrt{2\pi}\,\sqrt{\sigma_k^2+\sigma_n^2}} \,
    \exp\left( -\frac{\left(\mu_n-\mu_k\right)^2}{2(\sigma_k^2+\sigma_n^2)} \right) \,,
\end{align}
and again taking the product over
data points $\{\mu_n,\sigma_n\}$.

To fit \gmm[s] with
\mbox{$\Ncomp=1,2,3$} components
to the $\Mtot$ data set,
we can use
two independent \python packages:
\begin{enumerate}
 \item \sw{sklearn.mixture.GaussianMixture}~\citep[\skgmm for short,][]{Pedregosa:2012toh}
       supports basic multi-component \gmm fitting
       without measurement errors.
 \item \xdgmm~\citep[Xtreme Deconvolution GMM,][]{Holoien:2017xdg,xdgmm}
       can also handle known measurement errors;
       it serves as a wrapper for the \sw{astroML}~\citep{VanderPlas:2014wua}
       implementation of a method by~\citet{Bovy:2009ju}.
\end{enumerate}

Fit results are collected in Table~\ref{tab:DNS_Mt_stats}
and compared with those from \huang.
The main results of interest are those from \xdgmm
under consideration of measurement errors
(\emph{heteroscedastic} case).
These agree well with \huang
for one- and two-component \gmm[s],
with the small differences consistent
with different fitting implementations.
The likelihood ratio of
\mbox{$\likeli_1/\likeli_2\approx0.014$}
is also similar to their reported 0.011,
though going to three components
provides another factor of
\mbox{$\likeli_2/\likeli_3\approx0.03$}
which is less strong,
but already illustrates the danger
of model selection by likelihood ratios alone:
Adding additional components to the mixture model
will generally increase the likelihood
until each data point is fit by its own model component.
\huang estimate significance assuming a $\chi^2$ distribution for the likelihood ratio,
which is itself problematic for a \gmm on a small data set~\citep[see e.g.][]{Ciuperca:2003pen,Chen:2008inf,Chen:2009hyp};
instead the following sections will describe more robust hypothesis tests.

As a sanity check,
\xdgmm with errors set to zero
produces the same
best-fitting \gmm parameters and likelihoods
as \skgmm.
Using uniform errors
(\emph{homoscedastic} case)
is equivalent to no errors,
besides reducing the estimated $\sigma_k$,
as expected from Eq.~\ref{eq:gmm_errs}.
Parameter estimates are consistent in all three cases,
though likelihoods are different
and the full error treatment is important in assessing statistical robustness,
as we will see through the following series of tests.

\begin{table*}
 \centering
 \caption{\gmm results
          for the total masses $\Mtot$ of \numdns galactic \dns systems
          from~\citehuang.
          \skgmm and \xdgmm results without measurement errors
          are identical to the quoted precision.
          Quantities are defined in sections~\ref{sec:gmms}--\ref{sec:em-test}.
         }
 \label{tab:DNS_Mt_stats}
 \tabcolsep=0.1cm
 \resizebox{\textwidth}{!}{
  \hspace{-0.75cm}
  \begin{tabular}{ccccccccccccrrc} 
   \hline
    & $\Ncomp$ & $C_1$ & $\mu_1$ & $\sigma_1$ & $C_2$ & $\mu_2$ & $\sigma_2$ & $C_3$ & $\mu_3$ & $\sigma_3$ & $\log_{10}\mathcal{L}$ & $\AICc$ & $\BIC$ & $\logevi$ \\
\hline
\huang          & 1 & 1.00 & 2.67 & 0.10 &      &      &      &      &      &      & 5.77 \\
                & 2 & 0.40 & 2.58 & 0.01 & 0.60 & 2.72 & 0.08 &      &      &      & 7.77 \\
\hline
\skgmm          & 1 & 1.00 & 2.66 & 0.10 &      &      &      &      &      &      & 5.80 & -21.71 & -21.29 \\
\textit{or} \xdgmm & 2 & 0.51 & 2.58 & 0.02 & 0.49 & 2.75 & 0.07 &      &      &      & 8.15 & -20.88 & -24.00 \\
(no errors)     & 3 & 0.53 & 2.58 & 0.02 & 0.33 & 2.72 & 0.02 & 0.13 & 2.85 & 0.02 & 9.67 & -4.54 & -22.88 \\
\hline
\xdgmm          & 1 & 1.00 & 2.66 & 0.10 &      &      &      &      &      &      & 5.74 & -21.45 & -21.04 \\
(heterosc.)     & 2 & 0.48 & 2.58 & 0.02 & 0.52 & 2.74 & 0.07 &      &      &      & 7.60 & -18.35 & -21.48 \\
                & 3 & 0.52 & 2.58 & 0.02 & 0.35 & 2.71 & 0.02 & 0.13 & 2.85 & 0.02 & 9.07 & -1.79 & -20.13 \\
\hline
\cpnest         & 1 & $1.00_{-0.00}^{+0.00}$ & $2.67_{-0.04}^{+0.04}$ & $0.10_{-0.02}^{+0.03}$ &      &      &      &      &      &      &       &       &       & 8.32$\pm$0.06 \\[2pt]
                & 2 & $0.42_{-0.27}^{+0.42}$ & $2.58_{-0.01}^{+0.07}$ & $0.02_{-0.02}^{+0.09}$ & $0.58_{-0.42}^{+0.27}$ & $2.72_{-0.06}^{+0.08}$ & $0.09_{-0.04}^{+0.05}$ &      &      &      &       &       &       & 8.63$\pm$0.08 \\[2pt]
                & 3 & $0.38_{-0.38}^{+0.49}$ & $2.57_{-0.18}^{+0.03}$ & $0.02_{-0.02}^{+0.12}$ & $0.35_{-0.27}^{+0.32}$ & $2.68_{-0.10}^{+0.06}$ & $0.05_{-0.05}^{+0.09}$ & $0.26_{-0.22}^{+0.37}$ & $2.76_{-0.07}^{+0.12}$ & $0.08_{-0.07}^{+0.10}$ &       &       &       & 8.51$\pm$0.08 \\
\hline
\mixinf         & 1 & 1.00 & 2.66 & 0.10 &      &      &      \\
(no errors)     & 2 & 0.51 & 2.58 & 0.03 & 0.49 & 2.74 & 0.08 \\

   \hline
  \end{tabular}
 }
\end{table*}

\subsection{Information criteria: AICc and BIC}
\label{sec:aicbic}

In general,
when adding additional components to a \gmm
the model likelihood will keep increasing.
Hence, this test alone
can tempt into overfitting any given data set.
A more robust way of model selection
is provided by \emph{information criteria}
which introduce a penalty term for higher
numbers $\Ncoef$ of coefficients.
An astronomy-focused review and pedagogical introduction
to such information criteria
is provided by \citet{Liddle:2007fy}.
See also \citet{Burnham-Anderson:2002}
for a more in-depth exposition.

The Akaike Information Criterion (AIC),
originally introduced by \citet{Akaike:1974},
is given
in its modified form~\citep[the AICc,][]{Hurvich:1989} as
\begin{equation}
 \label{eq:AICc}
 \AICc = -2 \logL + 2\,\Ncoef + \frac{2\,\Ncoef\,(\Ncoef+1)}{\Ndata-\Ncoef-1} \,.
\end{equation}
Here the second term is the original Akaike penalty for complex models,
and the third term is a correction
to produce more reliable rankings when
$\Ndata$ is small.
(The AICc converges to the original AIC for large $\Ndata$.)

A popular alternative is
the Bayesian Information Citerion (BIC) introduced by \citet{Schwarz:1978}:
\begin{equation}
 \label{eq:BIC}
 \BIC = -2 \logL + \Ncoef \, \ln\Ndata \,.
\end{equation}
Despite its name,
it is in general not equivalent
to a full Bayesian evidence comparison between two models.

\emph{Lower} values of either criterion
indicate a preferred model
with a better balance between goodness-of-fit and parsimony.
The strength of preference is given purely
by the \textit{differences} between models:
any overall additive constant can be ignored.
There is no universal agreement on how large a difference
constitutes clear preference between models,
though values between 3 and 5 are usually
quoted~\citep{Raftery:1995bay,Burnham:2004mul,Liddle:2007fy}.
Note also that these criteria are formally motivated by asymptotic considerations
\citep[see e.g.][ and references therein]{Burnham-Anderson:2002,Burnham:2004mul}
which cannot be invoked for the small-$\Ndata$ problem under consideration here.
Hence, for now let us consider them as heuristic criteria,
and investigate how they compare with other tests.
(See also appendix~\ref{sec:appendix-sims}
for simulations with larger $\Ndata$.)

\begin{figure}
 \includegraphics[width=\columnwidth]{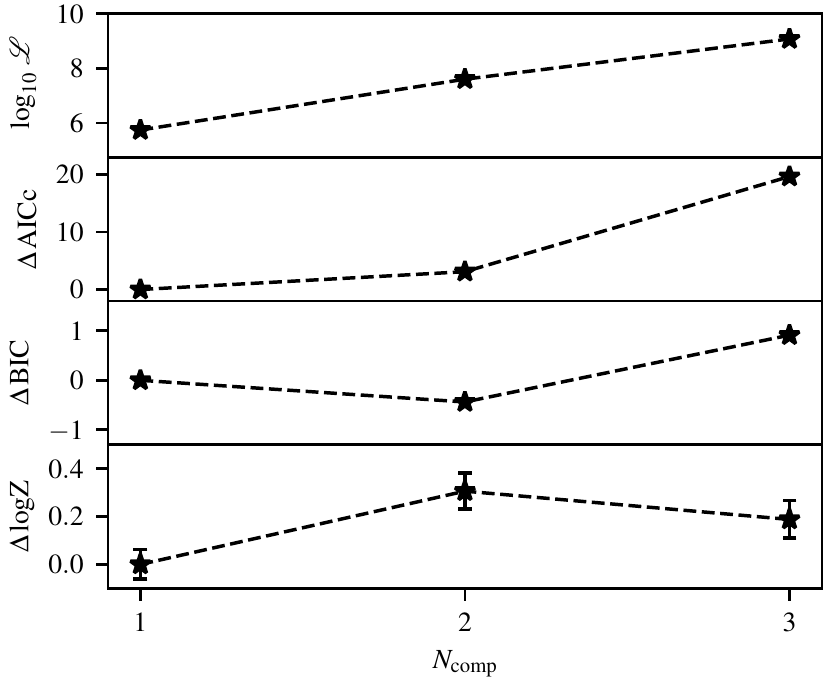}
 \vspace{-\baselineskip}
 \caption{Model selection criteria for \gmm[s]
          with \mbox{$\Ncomp=1,2,3$}
          applied to
          the \dns $\Mtot$ data set.
          The first three panels are for heteroscedastic \xdgmm fits.
          The last panel gives the average Bayesian evidence
          from 10 \cpnest runs at each $\Ncomp$ (Sec.~\ref{sec:nestsamp}).
          AICc, BIC and $\logevi$
          are plotted as differences $\Delta$
          to the \mbox{$\Ncomp=1$} value
          (e.g. \mbox{$\Delta\BIC(\Ncomp=2)=\BIC(\Ncomp=2)-\BIC(\Ncomp=1)$}).
          For AICc and BIC,
          \emph{negative} $\Delta$
          would mean a preference for that $\Ncomp$.
          \vspace{-\baselineskip}
         }
 \label{fig:DNS_Mt_criteria}
\end{figure}

Revisiting the heteroscedastic \xdgmm fits for the $\Mtot$ data set
using these three criteria,
Fig.~\ref{fig:DNS_Mt_criteria}
provides a comparison against the simple log-likelihood,
as a function of $\Ncomp$.
The penalty of the AICc is strong
for the present case of small $\Ndata$,
so that despite the likelihood ratio
it slightly prefers a single component
(by $\Delta\AICc\approx3$)
and very strongly rejects a third component.
The BIC gives very small differences,
telling us that the data are indecisive.
From Table~\ref{tab:DNS_Mt_stats},
note also that the no-errors fits
give a lower BIC for \mbox{$\Ncomp=2$},
and hence indeed the full error treatment
is important in obtaining a robust model selection --
the difference is easily understood by the clustering
of wide-uncertainty measurements near lower $\Mtot$.

Overall, these criteria
(unsurprisingly)
agree rather clearly that there is no justification
for adding a third \gmm component.
However for the main question of \huang,
whether there are two components or only one,
the situation is still indecisive.
As we will see from the alternative examples
in the appendix,
information criteria are generally expected
to converge on a consistent answer
when the data are indeed informative about the model selection question.
Hence, it appears that for
the $\Mtot$ distribution of Galactic \dns systems,
the data set is simply not yet large
(and/or precise)
enough to conclusively answer the question.

\subsection{Cross-validation}
\label{sec:crossval}

Another independent check
for overfitting is \cv.
The basic idea is to check
the intra-sample variance of a data set
by re-evaluating fits on subsets of the data.
For each iteration,
a figure of merit
(e.g. log-likelihood)
is computed on the left-out data points,
and in the end averaged over iterations.
(In other words,
for each iteration,
the left-out data are a `test' set
for a model `trained' on the remaining data.)
Overly complex models are expected
to get over-fit to the training subsets and then provide
inferior prediction performance on the test subsets.
The conceptually simplest version
is \emph{leave-one-out \cv},
where all possible subsets of $\Ndata-1$ data points
are exhaustively evaluated.

\begin{figure*}
 \includegraphics[width=\linewidth]{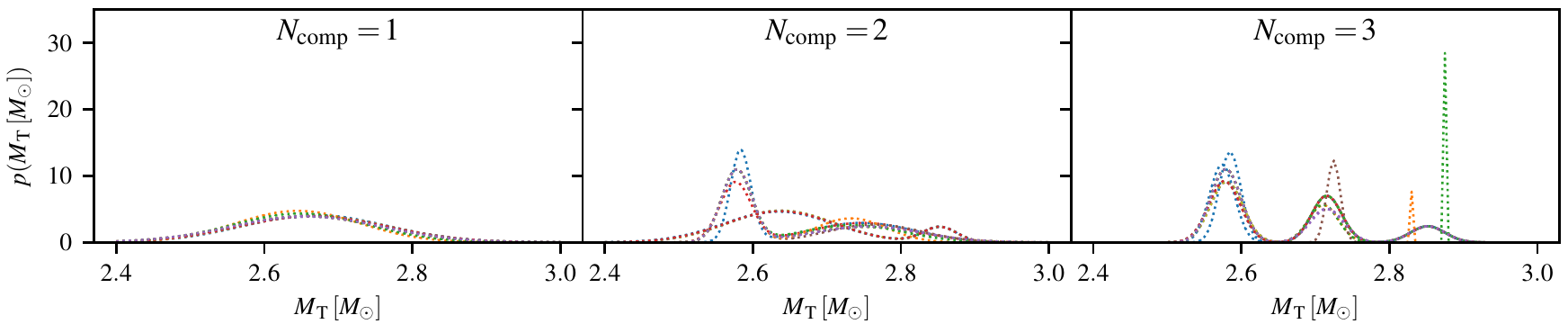} \\[0.5\baselineskip]
 \includegraphics[width=\linewidth]{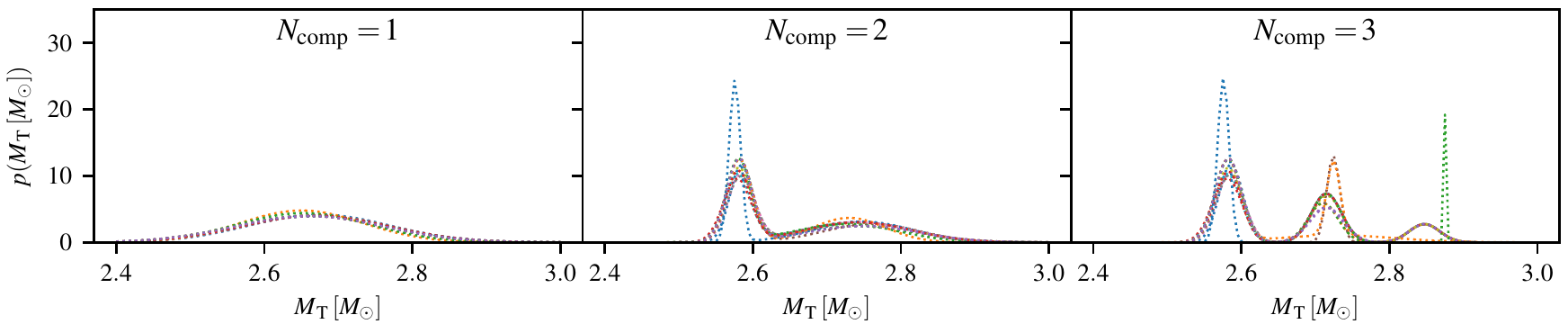}
 \vspace{-\baselineskip}
 \caption{Illustration of leave-one-out cross-validation
          on the $\Mtot$ data set.
          Top row: \skgmm fits
          ignoring individual measurement errors.
          Lower row: heteroscedastic \xdgmm fits.
          Each line corresponds to a fit
          after leaving out one data point.
         }
 \label{fig:DNS_Mt_crossval}
\end{figure*}

Numerical cross-validation scores
turn out not to be useful for this small data set,
as the variance is too large
to make any robust statements.
However, an illustrative analysis
in the spirit of leave-one-out \cv
is easily done by fitting \gmm[s] for
all \numdns subsets of 14 data points each.
This also helps identify
systems that have a large effect
on the fit.

The individual fitted distributions
for each iteration
are compared in Fig.~\ref{fig:DNS_Mt_crossval}.
When ignoring measurement errors,
individual systems in the
\mbox{$\Mtot\lesssim2.65\Msun$} range
have a large influence on the two-component fits,
with the lower-mass peak
sometimes even shifting
to within the visually apparent `gap'.
By contrast, in heteroscedastic fits,
some of those systems are already downweighted
by their large uncertainties,
and the leave-one-out fits become somewhat more stable.
Three-component fits are very unstable in either case.

Hence, this graphical version of a leave-one-out \cv test
supports a single-Gaussian fit as stable over data subsets,
and clearly cautions against three components.
Once measurement errors are taken into account,
this approach does not uncover any clearly apparent problems with
the two-component fit suggested by \huang,
but it is slightly less stable than one component.

\begin{figure*}
 \includegraphics[width=\textwidth]{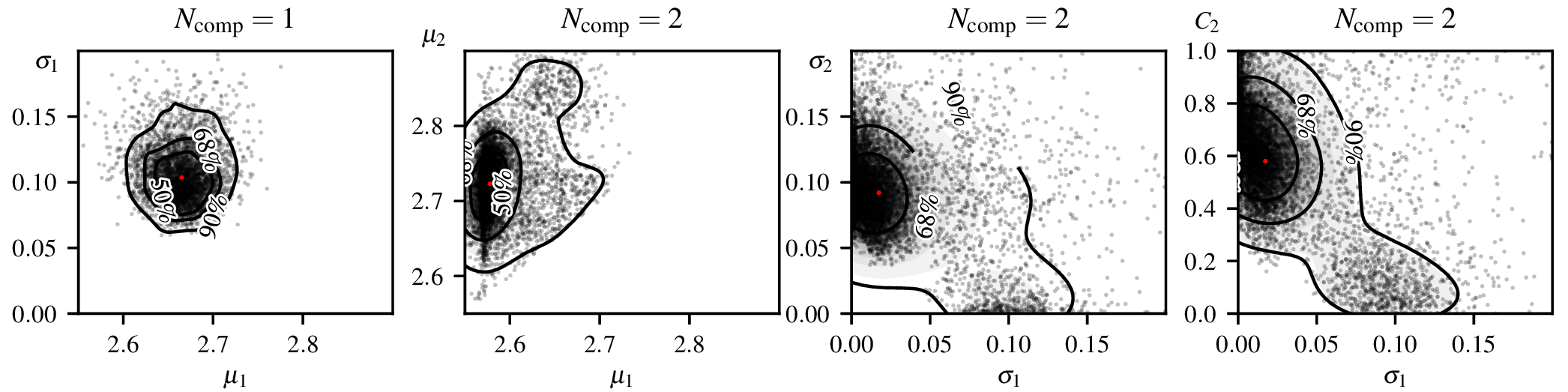}
 \vspace{-\baselineskip}
 \caption{Posteriors from \cpnest runs
          for a single Gaussian (left panel)
          and a two-component \gmm (remaining panels).
          Note that $\sigma_1$ appears twice
          on the x-axis of the last two panels,
          to better illustrate the posterior structures
          discussed in detail in appendix~\ref{sec:appendix-cpnest}.
          \vspace{-0.5\baselineskip}
         }
 \label{fig:DNS_Mt_post}
\end{figure*}

\subsection{Bayesian evidence from Nested Sampling}
\label{sec:nestsamp}

Since the \dns data set is so small
(\mbox{$\Ndata=\numdns$}),
it is computationally cheap to obtain
Bayesian posterior estimates
and evidences for model selection.
Starting from some prior knowledge $\pInfo$,
a prior distribution $\prob{\theta}{\,\Hyp,\pInfo}$
for the parameters $\theta$ of a model $\Hyp(\theta)$,
and the \gmm likelihood
\mbox{$\prob{x}{\theta,\Hyp,\pInfo} = \likeli \left( x_n\,|\,\{\mu_k,\sigma_k,C_k\} \right)$}
from Eq.~\ref{eq:gmm} or~\ref{eq:gmm_errs},
the posterior distribution for $\theta$ under that model follows from Bayes' theorem:
\begin{equation}
 \label{eq:bayes}
 \prob{\theta}{x,\Hyp,\pInfo} = \frac{\prob{\theta}{\,\Hyp,\pInfo}\prob{x}{\theta,\Hyp,\pInfo}}{\prob{x}{\,\Hyp,\pInfo}} \,.
\end{equation}
The Bayesian evidence for a model $\Hyp$ is defined as its likelihood
marginalised over its whole prior support,
\begin{equation}
 \label{eq:evidence}
 \evi_\Hyp = \prob{x}{\,\Hyp,\pInfo}
      = \int \mathrm{d}\theta \, \prob{\theta}{\,\Hyp,\pInfo} \, \prob{x}{\theta,\Hyp,\pInfo} \,.
\end{equation}
Note that this is still dependent on the model $\Hyp$,
whereas the total evidence $\prob{x}{\pInfo}$ would be a model-independent normalisation factor.
Evidence ratios, also called Bayes factors,
are a convenient quantity for model selection,
as priors need to be defined only over the parameter space of each model,
but not between models.
See \citet{Gregory:2005blda,Liddle:2007fy,Heavens:2009nx,jaynes2003:_logic} and references therein
for the underlying theory.

To evaluate $\evi_\Hyp$ for \gmm[s] of different $\Ncomp$,
we can use \cpnest~\citep{CPNest},
a \python implementation of the nested sampling algorithm by \citet{Skilling:2004ns},
with the heteroscedastic likelihood function (Eq.~\ref{eq:gmm_errs})
and \mbox{$\Nlive=1024$} sampler live points.
The code also provides an estimate for the uncertainty on the evidence,
$\Delta\logevi_\Hyp\approx\sqrt{H_\Hyp/\Nlive}$,
with the information gain $H_\Hyp$ from prior to posterior.

The outcome of Bayesian inference in general depends
on the choice of priors $\prob{\theta}{\,\Hyp,\pInfo}$;
the following results are obtained from weakly informative priors
which are discussed in detail in appendix~\ref{sec:appendix-cpnest}
along with a test for robustness under different choices.
Overall, the \cpnest posterior estimates and evidence ratios
appear stable under reasonable prior changes.

\cpnest results are also included in Table~\ref{tab:DNS_Mt_stats}.
The listed parameter estimates are posterior medians $\pm$
10\% and 90\% quantiles.
While these consistently include the previous estimates,
it is interesting to note that for \mbox{$\Ncomp=2$}
the posterior uncertainties on $\mu_k$ and $\sigma_k$
are also almost compatible with a vanishing separation between the two components,
and those on the $\sigma_k$ and component weights $C_k$ are rather large.
The \mbox{$\Ncomp=3$} case is not well constrained and hence posterior estimates are very broad,
with strongly overlapping components.
The \mbox{$\Ncomp=1,2$} posteriors are also illustrated in Fig.~\ref{fig:DNS_Mt_post}
and discussed in detail in appendix~\ref{sec:appendix-cpnest}.
In addition, Fig.~\ref{fig:DNS_Mt_cpnest_dist}
shows the median reconstructed \gmm distribution functions and their 90\% intervals.

No \cpnest likelihood point estimates are included in Table~\ref{tab:DNS_Mt_stats}
since these might be misleading without context:
Near the posterior median, $\logtenL$ is generally close to
the previous fit results,
while higher values can be found in some overall less favoured parts of parameter space.
(See appendix~\ref{sec:appendix-cpnest}.)
The main quantity of interest for model comparison,
the model evidence $\evi_\Hyp$,
is not derived from a point estimate,
but as seen in Eq.~\ref{eq:evidence} it takes into account
the whole sampled volume.
At
\mbox{$\evi_2/\evi_1\lesssim1.4$},
\mbox{$\evi_2/\evi_3\approx1.1$},
the evidence ratios are indecisive,
meaning that the increased prior volume of \gmm[s] with higher $\Ncomp$
just about makes up for the higher likelihoods achieved,
and no clear preference for either model can be found.
The alternative prior choices considered in appendix~\ref{sec:appendix-cpnest}
do not change $\evi_2/\evi_1$ far away from unity,
indicating that much tighter priors would be needed
to obtain clear preference for a multi-component model,
which would then however be driven by that prior choice and not by the data.

\begin{figure}
 \includegraphics[width=\columnwidth]{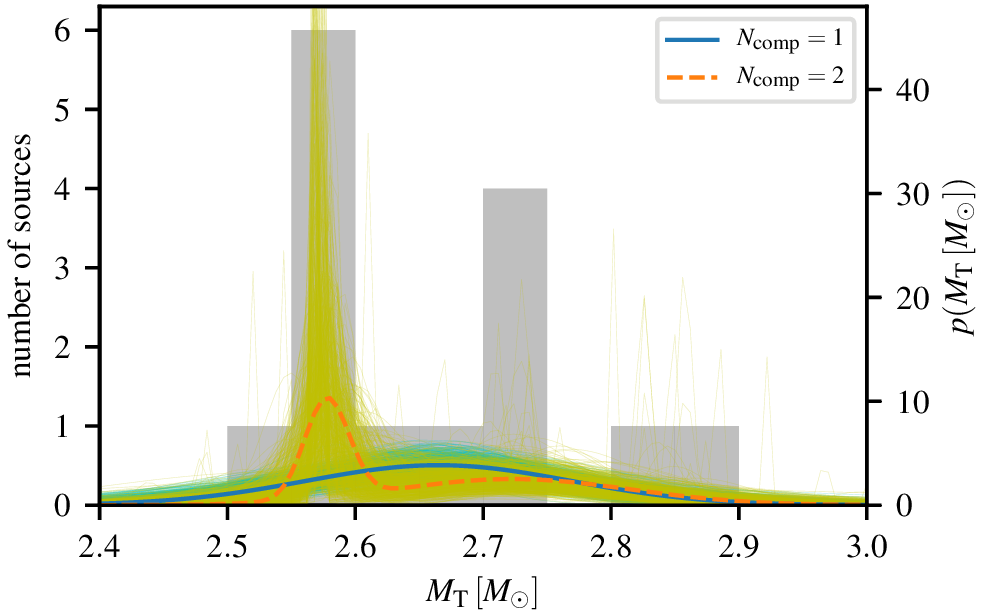}
 \vspace{-\baselineskip}
 \caption{\gmm distribution functions $p(\Mtot$)
          inferred with \cpnest for one (blue solid line)
          or two components (orange dashed line),
          superimposed on a data histogram.
          The solid foreground lines correspond to the median posterior parameters
          and the background `haze` is a superposition
          of $p(\Mtot$) evaluated at parameters within the 90\% confidence region.
          See appendix~\ref{sec:appendix-cpnest} for details.
         }
 \label{fig:DNS_Mt_cpnest_dist}
\end{figure}

\subsection{EM test}
\label{sec:em-test}

For finite mixture models, the EM-test \citep{Chen:2009hyp,Chen:2012inf}
is based on a penalised maximum-likelihood estimator (MLE)
and the expectation-maximisation (EM) algorithm~\citep{Dempster:1977max}.
An \sw{R} implementation is available in the package \mixinf~\citep{mixinf}.
In alternating iteration steps, EM assigns data samples to the proposed mixture components
according to their relative probability
and then the MLE is updated.~\citep[See][ for a didactic introduction.]{Do2008:exp}

\mixinf parameter estimates for \mbox{$\Ncomp=1,2$}
are included in Table~\ref{tab:DNS_Mt_stats}.
These are reasonably close to the \skgmm and \xdgmm results
and consistent with \cpnest within that method's uncertainty intervals.
For \mbox{$\Ncomp=3$}, \mixinf
returns 2 components identical to the \mbox{$\Ncomp=2$} model
and a completely negligible third, so this is not listed separately.

The EM-test statistic is a type of penalised likelihood.
The standard $p$-value assigned to the \mbox{$\Ncomp=1$} null hypothesis
by \mixinf is computed under an asymptotic large-$\Ndata$ assumption; 
nominally this returns \mbox{$p=0.087$} for the \dns $\Mtot$ data set
but due to the low \mbox{$\Ndata=\numdns$} this should be interpreted carefully.
A better understanding of the actual hypothesis test power
can be achieved through repeating the EM-test on simulated data,
see appendix \ref{sec:appendix-sims},
indicating that at a nominal $p$-value threshold (e.g. \mbox{$p=0.05$})
this test rejects too few data sets generated with \mbox{$\Ncomp=1$}.
Still, compared with those simulations,
the obtained EM-test result does not allow a confident rejection of the \mbox{$\Ncomp=1$} hypothesis either.
Another caveat is that \mixinf by default does not include measurement uncertainties.


\section{Conclusions}

The distribution of total masses $\Mtot$ of Galactic \dns systems
shows some apparent bimodality,
which can be fit with a two-component \gmm as shown by \huang.
A pure likelihood ratio test prefers those two components over one,
with \huang estimating the significance of this preference as $2\sigma$.
As a first step towards testing if this indeed points to
two distinct underlying populations of astrophysical objects,
while it is my understanding that \huang are also working on a more sophisticated analysis,
in this note I have kept their initial \gmm assumption,
but considered more robust model selection criteria:
Neither the frequentist information criteria (AICc and BIC)
considered in Sec.~\ref{sec:aicbic},
which amend the likelihood ratio test with a penalty
for the higher number of free parameters in multi-component \gmm[s],
nor a Bayesian evidence ratio test (Sec.~\ref{sec:nestsamp})
find any robust preference
for more than one component.
The various \gmm fitting methods employed here
still all agree with \huang
that a two-component \gmm certainly
provides `a good fit' to the data;
the scenario is not ruled out either
and, as pointed out by \huang,
could have interesting consequences for stellar evolution models and \gw astronomy.
But it appears that the present set of known \dns[s]
is simply too small,
and some systems' masses are not constrained well enough,
to robustly decide between one or two components.

It will be interesting to revisit this model selection problem
once additional \dns systems are observed,
as expected in great numbers from upcoming surveys
e.g. with MeerKAT~\citep{Bailes:2018azh}
and the SKA~\citep{Smits:2008cf};
or to combine the Galactic sample
with \gw observations of extragalactic mergers,
as suggested by \huang in the second half of their paper.
(Though~\citet{Pankow:2018iab} suggests that
GW170817~\citep{TheLIGOScientific:2017qsa} might not be consistent
with the same population
as the galactic \dns[s].)

In the meantime,
the simple reanalysis in this note
has certainly not exhausted the full potential
of the present data set.
One could also consider distribution functions
beyond the \gmm family.
(\huang already suggested a \gmm plus uniform distribution.)
And since the cross-validation analysis suggests that,
for the current small data set size,
a few systems can have a large effect on any inference of the underlying distribution,
revisiting individual systems' mass measurements --
or even their identity as \dns[s] --
could also improve the situation.
For example, \mbox{J1811--1736}
has the widest uncertainty
in the \huang data set
(\mbox{$\Mtot=2.57\pm0.10$});
referring back to the original studies of
\citet{Lyne:1999hu} and \citet{Corongiu:2006rd},
its rather low companion mass means
that while it is generally accepted as a \dns,
this identification might not be completely iron-clad.
A combined reanalysis of total and component masses
could also be promising in constraining the model selection problem,
and more sophisticated statistical techniques could be applied
to deal with possible selection effects.

Data sets used in this note
(reproduced from \citehuang and \citet{Patruno:2017oum})
and \cpnest posterior samples
are provided as \href{https://arxiv.org/src/1808.01129/anc}{ancillary files}
of the arXiv preprint.

\vspace{-\baselineskip}

\section*{Acknowledgements}

The author was funded under the EU Horizon2020 framework,
Marie Sk\l{}odowska-Curie
grant agreement 704094 GRANITE.
Thanks to Zhu Xingjiang and Paul Lasky (P.L.)
for inspirational discussions on the $\Mtot$ topic;
to both of them, to John Veitch (J.V.) and Graham Woan
for comments on the manuscript;
to J.V. for an introduction to \cpnest;
to P.L. for suggesting Fig. \ref{fig:DNS_Mt_cpnest_dist};
and to the anonymous MNRAS referee for suggesting the EM-test.




\bibliographystyle{mnras}
\bibliography{dns_mt}



\appendix

\section{GMMs on simulated DNS populations}
\label{sec:appendix-sims}

\begin{figure}
 \includegraphics[width=\columnwidth]{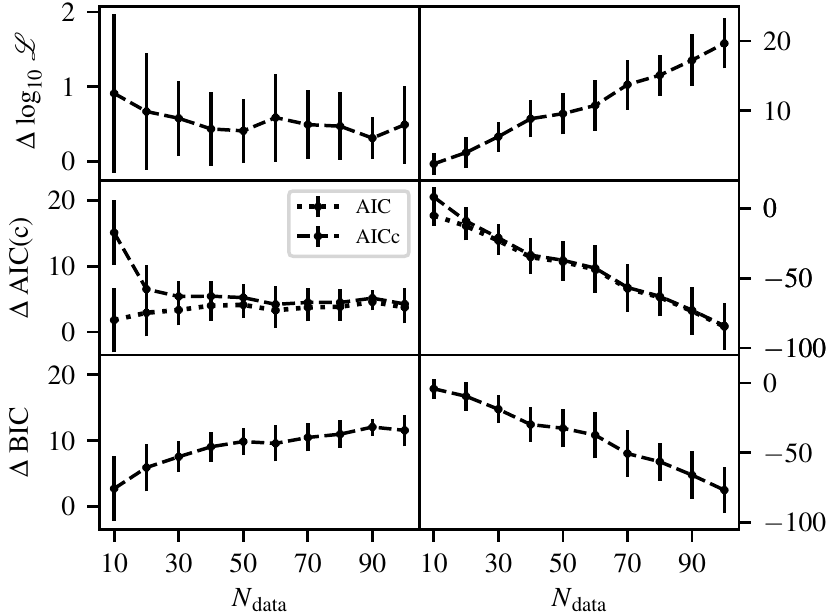}
 \vspace{-0.5\baselineskip}
 \caption{Model selection criteria for \gmm[s]
          with \mbox{$\Ncomp=1,2$}
          applied to
          simulated $\Mtot$ populations
          drawn from one- and two-component \gmm[s]
          with the fiducial values reported by \huang.
          50 independent populations are drawn
          for each $\Ndata$ step.
          Each difference $\Delta Q$ is taken as
          \mbox{$Q_{\Ncomp=2}-Q_{\Ncomp=1}$}.
          Hence, positive $\Delta\logtenL$
          imply a better fit of the two-component model,
          while for AIC and AICc
          (shown together in the second row)
          as well as BIC
          a \emph{negative} value indicates
          preference for the two-component \gmm.
          Error bars correspond to one standard deviation.
         }
 \label{fig:DNS_Mt_sims}
\end{figure}

To further illustrate the scaling and robustness
of various model selection criteria,
let us consider some simulated populations
where the `true' distribution is known.
Fig.~\ref{fig:DNS_Mt_sims} shows results
from simulated single Gaussians (left column)
or two-component mixtures (right column)
with parameters matching those reported by~\huang,
as a function of the number of randomly drawn samples $\Ndata$
in a data set.
For each step of 10 in $\Ndata$,
50 data sets are drawn,
and the differences in $\logtenL$, AIC, AICc and BIC
evaluated between one- and two-component models
fitted with \skgmm and without measurement errors.
(AIC is added here to demonstrate the convergence with AICc;
Bayesian evidences are omitted due to their higher computational cost.)

For a `true' single-component Gaussian,
the likelihood ratio test tends to stay inconclusive:
a two-component \gmm can always
fit the data slightly, but not much, better.
At small $\Ndata$, the AICc strongly prefers
the one-component model
due to its correction term,
while for large $\Ndata$
it converges to the AIC at $\Delta$ of $+3$ to $+5$,
and does not strengthen the case any further
due to the fixed complexity penalty of $2\,\Ncoef$.
On the other hand, the BIC continues to develop
a stronger preference for the `true' model
for increasing $\Ndata$
due to its different penalty term.

On the other hand, if the `true' model is a two-component \gmm,
all selection criteria agree in collecting
very strong preference ($\Delta<-10$) for it
by $\Ndata\gtrsim20$ to 30,
and continue to strengthen this preference
as $\Ndata$ increases.

Intuitively, this dichotomy makes sense:
For a `true' single Gaussian and even for large $\Ndata$,
a two-component \gmm fit can still always
approximate the observed distribution
by simply having the components overlap almost completely.
The likelihood is then almost the same,
and information criteria can only decide by their penalty term.
However, for draws from a two-component \gmm,
large $\Ndata$ will make any one-component fit
disagree strongly with the data,
and the fit improvement of two components
will easily make the $\logL$ contribution
dominate over any penalty term.

A similar set of simulations is also useful
to study the EM-test for low $\Ndata$,
putting the nominal results obtained in Sec.~\ref{sec:em-test}
into proper perspective in the non-asymptotic regime.
From 6000 simulations of \mbox{$\Ndata=15$} samples from `true` single-component $\Mtot$ distributions
with the nominal \huang parameters,
only 2.8\% produce $p$-values below the nominal 5\% threshold.
This indicates that, taking the $p$-value estimated from the asymptotic expressions at face value,
the null hypothesis would not be rejected often enough,
and hence the \mbox{$p=0.087$} result for the \dns data set cannot necessarily
be taken as failure to reject \mbox{$\Ncomp=1$}.
The empirical 95\% quantile of EM-test statistics from these simulations is
4.82, corresponding to an asymptotic $p$-value of 0.09,
so that the \dns result is suggestively close to this threshold.
However with \mbox{$p=0.05$} being a rather lenient threshold for the physical sciences to begin with,
and the caveat of the \mixinf EM-test implementation not considering measurement errors,
this borderline result cannot be confidently interpreted as evidence for \mbox{$\Ncomp>1$} either.

For comparison,
on simulations with \mbox{$\Ndata=15$} and `true' \mbox{$\Ncomp=2$},
an EM-test with chosen threshold of \mbox{$p=0.05$} on the asymptotic $p$-value
would reject the \mbox{$\Ncomp=1$} hypothesis in about 48\% of cases.
And for \mbox{$\Ndata=150$}, the empirical rejection rate indeed becomes
$\approx5\%$ for \mbox{$\Ncomp=1$} simulations
and $\approx100\%$ for \mbox{$\Ncomp=2$} simulations.

\section{Nested sampling: prior choice and additional details}
\label{sec:appendix-cpnest}

This section gives some additional details on the \cpnest runs
and interpretation of the resulting posteriors and evidences.

\vspace{0.5\baselineskip}

\textit{Prior choice:}
The \cpnest results in Sec.~\ref{sec:nestsamp}
use uniform priors in $[0,1]$ for the \gmm weights $C_k$,
truncated log-uniform priors in $[0.001,1]$ for the $\sigma_k$
and Gaussian priors for the $\mu_k$
with widths $0.3$ and means spaced uniformly in the range of $\Mtot$.
Constraints enforce
\mbox{$\sum_k C_k=1$}
and
\mbox{$\mu_{k+1} \geq \mu_k \forall k$}.

In Bayesian inference,
it is generally wise to test the effect of different prior choices.
For example, keeping the same priors on $\sigma_k$ and $C_k$
but changing the $\mu_k$ priors to uniform within $[1.8,4.0]\,\Msun$
yields almost unchanged posterior estimates for \mbox{$\Ncomp=1$}
and slightly broader, but consistent estimates for \mbox{$\Ncomp=2$}.
Alternatively, keeping the $\mu_k$ and $C_k$ priors
but narrowing the log-uniform range for the $\sigma_k$ to $[0.005,0.2]$
cuts off the third minor peak in the \mbox{$\Ncomp=2$} posteriors,
but only marginally influences the overall estimates.
The evidence values change somewhat with the prior volume,
but the ratios remain indecisive:
\mbox{$\evi_2/\evi_1\approx\exp(8.63-8.32)\approx1.4$} for the first set of priors,
\mbox{$\evi_2/\evi_1\approx\exp(6.28-7.04)\approx0.5$} for the uniform $\mu_k$ priors and
\mbox{$\evi_2/\evi_1\approx\exp(9.55-8.94)\approx1.8$} for the narrower $\sigma_k$ priors.

\vspace{0.5\baselineskip}

\textit{Posterior estimates:}
Let us consider the obtained posterior distributions a bit more closely,
especially the \mbox{$\Ncomp=1,2$} cases illustrated in Fig.~\ref{fig:DNS_Mt_post}.
The posterior parameter estimates given in Table~\ref{tab:DNS_Mt_stats}
are medians $\pm$ 90\% quantiles.
However, the \mbox{$\Ncomp=2$} posteriors have some asymmetric,
and for $\sigma_k$ even multimodal,
structure.
In general there are various arguments for or against quoting posterior medians vs. means,
see e.g. \citet{jaynes2003:_logic},
but for such cases medians tend to be more robust.
In any case, we can understand these features as not just due to technical issues,
e.g. insufficient convergence of the sampler
but to features of the underlying $\Mtot$ data set
and the fact that a single-component Gaussian is effectively included
in the paramer space of a two-component \gmm.
In the \mbox{$\Ncomp=2$} posteriors,
the smaller secondary peak at high $\sigma_1$ and low $\sigma_2$ corresponds
to very low $C_2$: in this part of parameter space, the posterior $p(\Mtot)$ \gmm function
is effectively almost a single, broad Gaussian with only a small localised bump at higher masses.
Conversely, there is a third, even smaller peak in the posterior
for low $\sigma_1$, high $\sigma_2$ and high $C_2$,
where the result is a dominant broad Gaussian with a small localised bump
near the observed low-$\Mtot$ excess.
These subdominant solutions with localised bumps can also be seen in some of the distribution functions
from within the 90\% quantiles plotted as background lines in Fig.~\ref{fig:DNS_Mt_cpnest_dist}.
These observations are consistent with the indecisive evidence ratio between \mbox{$\Ncomp=1$} and 2,
but they do also tell us to stay cautious of overfitting
even when additional data points will become available.

Along similar lines, for \mbox{$\Ncomp=3$}
the degeneracies become worse and the posteriors more complicated;
results in Table~\ref{tab:DNS_Mt_stats} are just quoted for completeness
and it is not particularly edifying to analyse the posteriors in detail.

\vspace{0.5\baselineskip}

\textit{Likelihoods:}
The nontrivial posterior structure for \mbox{$\Ncomp\geq2$}
leads to an ambiguity when trying to quote `the likelihood' from a nested-sampling run,
because in general neither the maximum-likelihood (ML) point nor the mode of the posterior (MAP)
need to be particularly close to where
the main mass of posterior probability is concentrated in parameter space.
For \mbox{$\Ncomp=1$}, both the ML and MAP agree with
the $\xdgmm$ results, and with
the likelihoods evaluated at the median or mean posterior parameter estimates,
to within $\logtenL\pm0.05$.
For \mbox{$\Ncomp\geq2$} the likelihoods at median or mean
still are similar to the $\xdgmm$ results,
while the ML and MAP values can be significantly higher (up to a factor of 10);
but these tend to come from extreme-$C_k$ parts of the parameter space
corresponding to the `dominant broad component plus small bump'
over-fitting cases discussed above.

\section{Comparison example: LMXB spin frequencies}
\label{sec:appendix-lmxbs}

As a comparative example,
consider the same \gmm analysis applied
to a completely different real life data set,
which shares the basic statistical properties
and model selection question with the \dns study at hand:
the distribution of spin frequencies $\fspin$ for a population
of 29 neutron stars in \lmxb systems.
The data set is given in Table 2 of~\citet{Patruno:2017oum}.
Those authors also fit one- and two-component \gmm[s]
(with an \sw{R} implementation of the EM algorithm)
and found a BIC difference of $\gtrsim7$ in preference of two components.

Ignoring measurement errors in this case
(which for \ns spin frequencies should be much smaller than for masses),
the \skgmm, \xdgmm and \mixinf implementations return consistent
parameters as reported in Table~\ref{tab:LMXB_fspin_stats}.
The \mbox{$\Ncomp=2$} results are fully consistent
with those reported by~\citet{Patruno:2017oum}
within their confidence intervals.

\begin{figure}
 \includegraphics[width=\columnwidth]{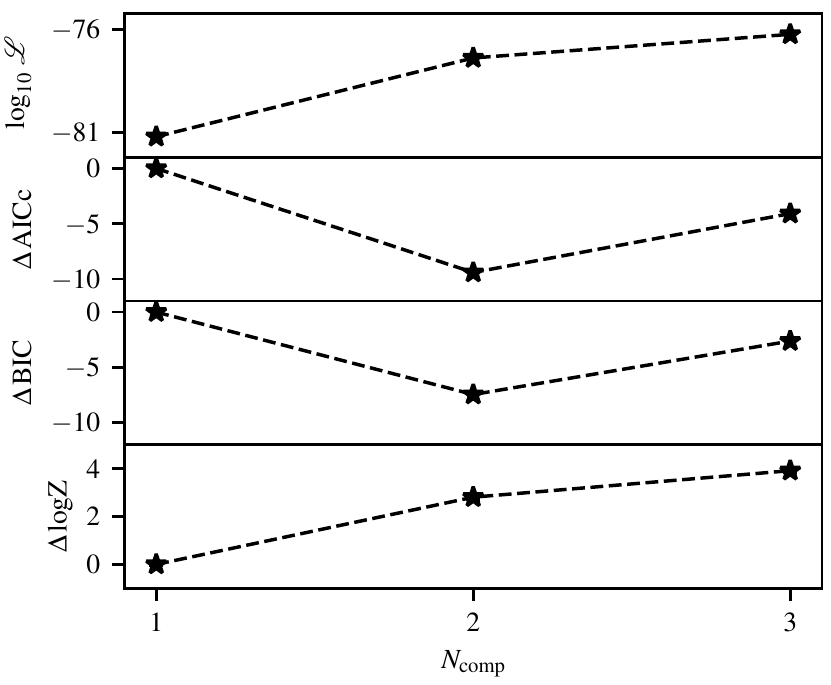}
 \vspace{-1.5\baselineskip}
 \caption{Model selection criteria for \gmm[s]
          with \mbox{$\Ncomp=1,2,3$}
          applied to
          the Patruno+ data set
          of $\fspin$ for \ns[s] in \lmxb[s].
          The first three panels are for \skgmm fits.
          The last panel gives the average Bayesian evidence
          from 10 \cpnest runs at each $\Ncomp$.
          AICc, BIC and $\logevi$
          are plotted as differences
          to the \mbox{$\Ncomp=1$} value.
          \vspace{-\baselineskip}
         }
 \label{fig:LMXB_fspin_criteria}
\end{figure}

\begin{figure*}
 \includegraphics[width=\linewidth]{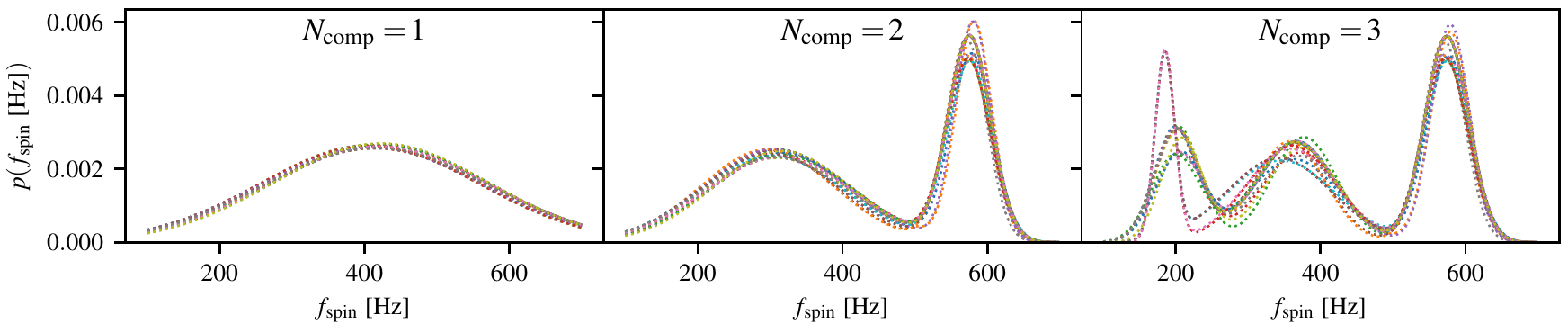}
 \vspace{-\baselineskip}
 \caption{Illustration of leave-one-out cross-validation
          on the Patruno+ $\fspin$ data set.
          Each dotted line corresponds to a fit
          after leaving out one data point.
          In this case,
          the fitted distributions are relatively stable
          not just for \mbox{$\Ncomp=1$},
          but also for the preferred fit with \mbox{$\Ncomp=2$},
          and only for even more components
          they start becoming unstable.
         }
 \label{fig:LMXB_fspin_crossval}
\end{figure*}

Looking at the various statistical criteria
as a function of number of components \mbox{$\Ncomp=1,2,3$}
(also illustrated in Fig.~\ref{fig:LMXB_fspin_criteria}),
these are also fully consistent with the findings of \citet{Patruno:2017oum}:
Again the likelihood ratio alone is already
in favour of two components
(\mbox{$\likeli_1/\likeli_2\approx10^{-4}$}),
but only additional criteria
can yield a robust decision,
and on its own,
the additional gain of
\mbox{$\likeli_2/\likeli_3\approx0.07$}
could tempt us to use an even more complex model.
Fortunately, in this case AICc and BIC
agree in clearly preferring two components over one
(with $\Delta$ of 7--11),
and also prefer two over three by $\Delta$ of almost 5.
A leave-one-out \cv test
(see Fig.~\ref{fig:LMXB_fspin_crossval})
also shows the two-component fit
to be almost as stable as a single component in this case.
A \mixinf EM-test reports a nominal $p$-value of 0.008
which also seems to reject a single-component hypothesis
much more clearly than in the DNS case.
As in Sec.~\ref{sec:em-test} the exact value needs to be interpreted with caution
due to the low $\Ndata$,
but this being only a comparison example,
no further simulation tests have been conducted in this case.

\cpnest results for this data set
(with uniform priors in $C_k$,
log-uniform in $\sigma_k$ within $[10,350]\,$Hz
and Gaussian $\mu_k$ priors of width 70\,Hz)
are very consistent with the other \gmm fits,
except for some differences in the more degenerate \mbox{$\Ncomp=3$} case.
The evidence ratio is in favour by $\geq16$ of two components over one,
with three components preferred by a marginal factor of only 3
and much less robust results.

Overall, the \lmxb data set seems to be a good example
for information criteria, EM-test and Bayesian evidence
consistently backing up
the selection of a more complex model:
in contrast to the $\Mtot$ of \dns[s] data set,
the preference for two components over a single Gaussian distribution
seems statistically robust.
Three components cannot be quite as confidently excluded,
but two yield the most robust fit.

\begin{table*}
 \centering
 \caption{\gmm results for the $\fspin$ data set
          for \ns[s] in \lmxb[s]
          from~\citet{Patruno:2017oum},
          assuming negligible measurement errors.
         }
 \label{tab:LMXB_fspin_stats}
 \tabcolsep=0.1cm
 \resizebox{\textwidth}{!}{
  \hspace{-0.75cm}
  \begin{tabular}{ccccccccccccrrc} 
   \hline
    & $\Ncomp$ & $C_1$ & $\mu_1$ & $\sigma_1$ & $C_2$ & $\mu_2$ & $\sigma_2$ & $C_3$ & $\mu_3$ & $\sigma_3$ & $\log_{10}\,\mathcal{L}$ & $\AICc$ & $\BIC$ & $\logevi$ \\
\hline
\skgmm             & 1 & 1.00 & 414 & 153 &      &     &     &      &     &     & -81.20 & 378.42 & 380.69 \\
\textit{or} \xdgmm & 2 & 0.60 & 308 &  99 & 0.40 & 576 &  29 &      &     &     & -77.39 & 369.00 & 373.23 \\
                   & 3 & 0.22 & 201 &  30 & 0.36 & 365 &  55 & 0.41 & 574 &  30 & -76.25 & 374.33 & 378.07 \\
\hline
\cpnest            & 1 & $1.00_{-0.00}^{+0.00}$ & $410_{- 34}^{+ 34}$ & $157_{- 24}^{+ 31}$ &      &     &     &      &     &     &       &       &       & -190.26$\pm$0.05 \\[2pt]
                   & 2 & $0.62_{-0.15}^{+0.14}$ & $315_{- 38}^{+ 42}$ & $110_{- 28}^{+ 39}$ & $0.38_{-0.14}^{+0.15}$ & $574_{- 25}^{+ 15}$ & $ 31_{- 11}^{+ 27}$ &      &     &     &       &       &       & -187.45$\pm$0.08 \\[2pt]
                   & 3 & $0.31_{-0.31}^{+0.40}$ & $255_{- 67}^{+ 64}$ & $ 76_{- 58}^{+ 62}$ & $0.33_{-0.27}^{+0.26}$ & $381_{- 50}^{+ 60}$ & $ 91_{- 59}^{+ 66}$ & $0.35_{-0.14}^{+0.14}$ & $577_{- 19}^{+ 15}$ & $ 30_{- 12}^{+ 19}$ &       &       &       & -186.35$\pm$0.08 \\
\hline
\mixinf            & 1 & 1.00 & 414 & 155 &      &     &     \\
                   & 2 & 0.60 & 307 & 100 & 0.40 & 575 &  41 \\

   \hline
  \end{tabular}
 }
\end{table*}

\begin{figure*}
 \includegraphics[width=\textwidth]{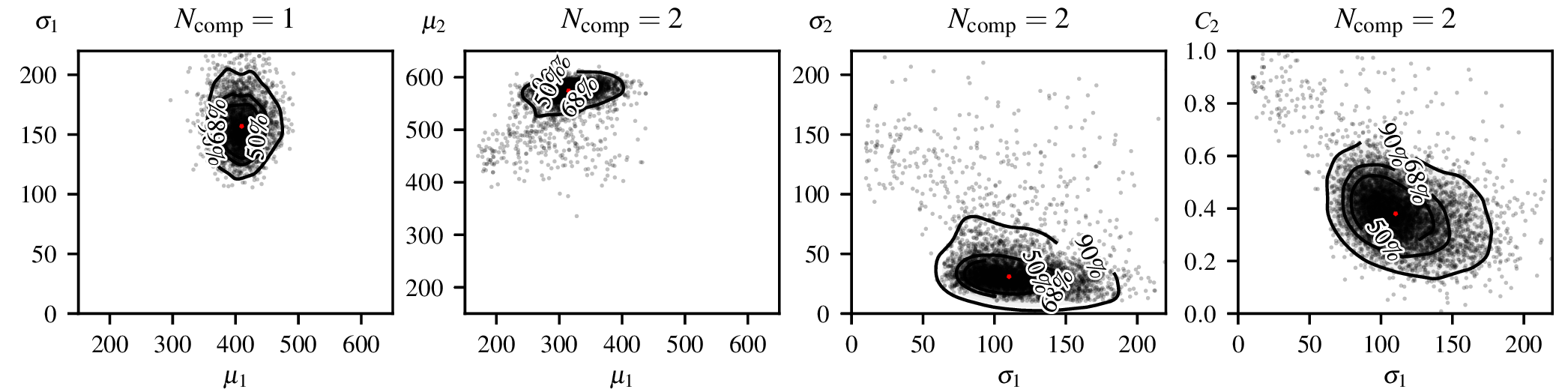}
 \vspace{-\baselineskip}
 \caption{Posteriors from \cpnest runs
          on the LMXB data set
          for a single Gaussian (left panel)
          and a two-component \gmm (remaining panels).
          Note that $\sigma_1$ appears twice
          on the x-axis of the last two panels.
          \vspace{-\baselineskip}
         }
 \label{fig:LMXB_fspin_post}
\end{figure*}

\begin{figure*}
 \begin{minipage}[c]{\columnwidth}
  \vspace{\baselineskip}
  \includegraphics[width=\columnwidth]{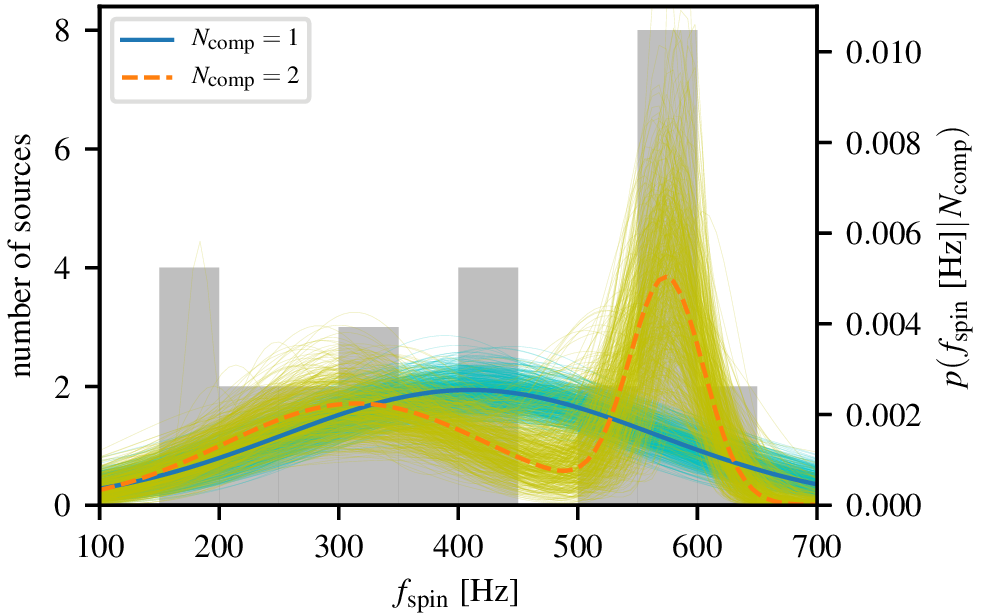}
 \end{minipage}
 \begin{minipage}[c]{\columnwidth}
 \caption{Median \gmm distribution functions $p(\fspin$)
          inferred with \cpnest for one (blue solid line)
          or two components (orange dashed line),
          superimposed on a data histogram,
          and with 90\% quantiles background `haze'.
         }
 \label{fig:LMXB_fspin_cpnest_dist}
 \end{minipage}
\end{figure*}

\bsp	
\label{lastpage}
\end{document}